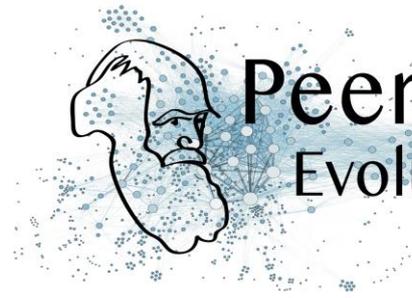

RESEARCH ARTICLE

# Testing host-plant driven speciation in phytophagous insects : a phylogenetic perspective

Emmanuelle Jousselin [1], Marianne Elias [2]

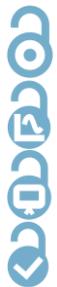

Open Access
Open Data
Open Code
Open Peer-Review

[1.] CBGP, INRA, CIRAD, IRD, Montpellier SupAgro, Univ Montpellier, Montpellier, France

[2] Institut Systématique Evolution Biodiversité (ISYEB), Muséum national d'Histoire naturelle, CNRS, Sorbonne Université, EPHE, Université des Antilles, 57 rue Cuvier, CP 50, 75005 Paris, France.

This article has been peer-reviewed and recommended by
*Peer Community in Evolutionary Biology*

**ABSTRACT**

During the last two decades, ecological speciation has been a major research theme in evolutionary biology. Ecological speciation occurs when reproductive isolation between populations evolves as a result of niche differentiation. Phytophagous insects represent model systems for the study of this evolutionary process. The host-plants on which these insects feed and often spend parts of their life cycle constitute ideal agents of divergent selection for these organisms. Adaptation to feeding on different host-plant species can potentially lead to ecological specialization of populations and subsequent speciation. This process is thought to have given birth to the astonishing diversity of phytophagous insects and is often put forward in macroevolutionary scenarios of insect diversification. Consequently, numerous phylogenetic studies on phytophagous insects have aimed at testing whether speciation driven by host-plant adaptation is the main pathway for the diversification of the groups under investigation. The increasing availability of comprehensive and well-resolved phylogenies and the recent developments in phylogenetic comparative methods are offering an unprecedented opportunity to test hypotheses on insect diversification at a macroevolutionary scale, in a robust phylogenetic framework. Our purpose here is to review the contribution of phylogenetic analyses to investigate the importance of plant-mediated speciation in the diversification of phytophagous insects and to present suggestions for future developments in this field.

# Introduction

The idea according to which new species arise through adaptation to different ecological niches constitutes the core of Darwin's work. This process is now termed ecological speciation and its study has become an intense field of research in evolutionary biology (Nosil 2012; Nosil *et al.* 2002; Rundle & Nosil 2005; Schluter 2009). **Phytophagous insects***[1] have always been at the forefront of these investigations (Drès & Mallet 2002; Elias *et al.* 2012; Forbes *et al.* 2017; Funk *et al.* 2002; Matsubayashi *et al.* 2010). The hypothesis of ecological speciation resulting from divergent selection exerted by host-plants was put forward a long time ago to explain the formation of new species of insects (Brues 1924; Walsh 1864). This scenario has been explored on several study systems. One text-book example of host-plant driven incipient speciation is the apple maggot (*Rhagoletis pomonella* complex) in which the evolution of new feeding preferences on the recently introduced domesticated apple (*Malus pumila*) has supposedly led to the emergence of specialized host races (Berlocher 2000; Bush 1975; Powell *et al.* 2014). Stick insects, leaf beetles (Nosil *et al.* 2012; Rundle *et al.* 2000), butterflies (McBride & Singer 2010), and the pea aphid also stand out as model systems in the study of host-driven speciation (Via *et al.* 2000; Caillaud & Via 2000; Peccoud *et al.* 2009; Smadja *et al.* 2012). In all these examples, the speciation scenario hypothesizes that: 1) the restricted utilization of distinct sets of host-plant species by insect populations is the result of adaptive trade-offs; 2) hybrids with intermediate phenotypes (in terms of traits involved in **host-plant adaptation***) fare poorly on parental host-plants and are selected against and therefore, gene flow between populations is reduced; 3) gene flow can further be reduced through the evolution of assortative mating, especially when host-plants also represent mating sites. In support of this scenario, many studies show the existence of genetically differentiated host races in insect species. Some studies have quantified selection against hybrids (McBride & Singer 2010; Gow *et al.* 2007) and some studies have uncovered genomic regions that determine host-plant preference and performance on alternative hosts (Egan *et al.* 2008; Smadja *et al.* 2012).

The role of host-plant-mediated speciation in the diversification of insects is also largely emphasized in the literature on large-scale patterns of insect diversity: macroevolutionary perspectives on phytophagous insect evolution have attributed their extraordinary diversification to selective responses to their host-plants (Ehrlich & Raven 1964; Janz 2011; Winkler & Mitter 2008; Yokoyama 1995). However, these macroevolutionary scenarios are often presented in the literature as narratives for specific insect lineages commenting on a phylogenetic reconstruction of the history of host-plant associations. Many phylogenetic studies still fail to clearly formulate hypotheses and predictions about the speciation processes that underlie the observed patterns and the role played by host-plant adaptation in those. The reason might be that the macroevolutionary patterns that arise when **host-plant specialization*** is the driver of speciation events are not always clear. There is no review on what to expect and how to formally test these predictions.

The increasing availability of robust molecular phylogenies and recent developments in phylogenetic comparative methods are offering an unprecedented opportunity to test evolutionary hypotheses in a robust phylogenetic framework. Our purpose here is to present the macroevolutionary scenarios for the diversification of phytophagous insects that have been put forward in the literature, decipher the role that ecological speciation driven by host-plant adaptation plays in them and synthetize predictions from these scenarios. We then identify tools from the "comparative phylogenetic toolbox" that provide ways to test some of these predictions. This toolbox can be divided into three compartments.

---

[1] See glossary

1) Comparisons of the phylogenetic histories of insects and their associated plants: the congruence (in terms of dates of divergence and branching patterns) of the phylogenetic histories of plant-feeding insects and their host-plants can be tested in robust statistical frameworks and illuminate how herbivores track the diversification of their hosts.

2) Ancestral character state reconstructions: the evolutionary trajectory of host-associations, host breadth and host-plant adapted traits can be inferred using ancestral character state reconstructions. Statistical tests can then determine whether their distribution throughout the phylogenetic trees follow the predictions of scenarios involving ecological speciation mediated by host-plant adaptation.

3) Diversification analyses: the recent developments of methods to study the **diversification dynamics**\* of entire clades using phylogenetic trees provides ways to test how shifts to new host-plant species or changes in host breadth have impacted diversification rates in phytophagous insects.

We review papers that have adopted these approaches. We then present suggestions for future research that should help linking microevolutionary studies on host-plant adaptation and macroevolutionary perspectives on phytophagous insect diversification.

### I Macroevolutionary scenarios of phytophagous insect diversification

#### I.1) *Escape and radiate* (Figure 1 a)

More than 50 years ago, Ehrlich & Raven (1964) put forward a macroevolutionary scenario that inspired most of the current research on plant-feeding insect diversification: it is known as "***Escape and radiate***" (Thompson 1989). They hypothesized that when insects acquire the ability to circumvent the chemical defenses of a plant group, it promotes their rapid diversification by ecological release, *i.e.* the availability of novel resources and reduction in direct competition. Insects undergo an **adaptive radiation**\*. In this scenario, adaptation towards host-plants is therefore the driving force of insect species formation. The *Escape and Radiate* scenario also hypothesizes that, in response to phytophagous insect predation, plants acquire novel chemical defenses which allow them in turn to diversify very rapidly (Marquis *et al.* 2016). Ehrlich & Raven's seminal study suffers from several shortcomings that have been pinpointed before (Janz 2011). First, although the authors frame their theory within the concept of adaptive radiation, they do not explicitly lay out any speciation mechanisms for the partners of the interaction. Following their scenario, a trade-off in resource use and specialization towards specific host-plants is necessary to explain the formation of numerous insect species (*i.e.,* species radiation) after the capture of a new host-plant lineage. Such a trade-off is not mentioned in the original paper (Janz 2011). In addition, as underlined by contemporary researchers of Ehrlich and Raven, it is difficult to conceive how the selection pressures exerted by insects on plant defences can drive plant speciation (Jermy 1976;

Jermy 1984). Plant traits that reduce phytophagous insect attacks are rarely linked with reproductive isolation between plant populations (but see Marquis *et al.* 2016 for a review of scenarios of herbivore-induced speciation in plants) and the evidence for bursts of speciation in plants following the evolution of chemical defence is scant (Futuyma & Agrawal 2009). However this study has been and remains a great source of inspiration for studies on the diversification of plant-insect associations. This is probably because it is one of the first studies that attempts to explain how microevolutionary processes (host-plant adaptation) translate into macroevolutionary patterns (radiation onto newly acquired plant lineages). Several predictions that can be tested on phylogenetic trees have been derived from the *Escape and Radiate* scenario (Table 1).

In the years following its publication, *Escape and Radiate* was often interpreted as generating **cospeciation*** patterns; however it is now recognized that it rather predicts the sequential speciation of insects onto an already diversified plant lineage (Janz 2011; Suchan & Alvarez 2015). According to this prediction: 1) the reconstruction of the history of host-plant associations on the phylogenetic trees of insects should reveal **host-plant conservatism***, *i.e.* the use of related plant species by related insects (Winkler & Mitter 2008); 2) the phylogenies of herbivorous insects and their host-plants should be more congruent than expected by chance and the diversification of the insects should lag behind that of their host-plants; this is called sequential evolution (Jermy 1984) or host tracking. Nevertheless, when the association between insects and their host plants is species-specific, a pattern of cospeciation can be expected through simple co-vicariance: geographic barriers affect the differentiation of populations of interacting lineages in a similar way and cause simultaneous speciation events (Althoff *et al.* 2012; Brookes *et al.* 2015; Martínez-Aquino 2016). In these cases, it is geographic isolation and not natural selection that initiates the reproductive isolation of insect populations and subsequent speciation. However, the specificity of the insects and host-adapted traits enhance the probability of shared vicariant events.

According to the *Escape and Radiate* scenario, the diversification dynamics of insects should follow the typical pattern of adaptive radiations (Janz 2011), *i.e.* insects should show an acceleration of speciation rate upon the capture of new plant lineages or the evolution of detoxification mechanisms (Wheat *et al.* 2007) and then slow down when their niches are saturated (when species diversity is reaching the carrying capacity of the host-plant lineage) (Rabosky & Lovette 2008). Furthermore, the capture of a species-rich clade of plants should result in higher speciation rates than the capture of lineages encompassing less species (Roskam 1985).

### I.2) The *Oscillation Hypothesis* (Figure 1b)

The *Escape and Radiate* scenario was revisited more than a decade ago by Janz and collaborators (Janz & Nylin 2008; Janz *et al.* 2006; Nylin & Janz 2009).

Using butterflies as study systems, they stated that expansions in diet breadth followed by specialisation onto new host-plant species constantly fuel the diversification of phytophagous insects. This has been termed the "***Oscillation Hypothesis***" (Janz & Nylin 2008).

This hypothesis stipulates that transitions towards a generalist diet generally open up a new adaptive zone, which favours the capture of new host-plants. In this scenario, expansions in diet are enabled by the phenotypic plasticity of insects with respect to host-plants (Nylin & Janz 2009).

a) *Escape and Radiate* (Erlich & Raven 1964)

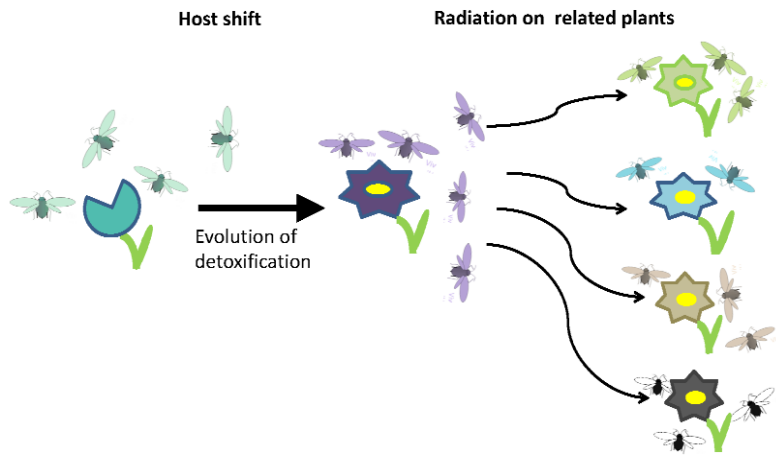

b) *Oscillation hypothesis* (Janz & Nylin 2008)

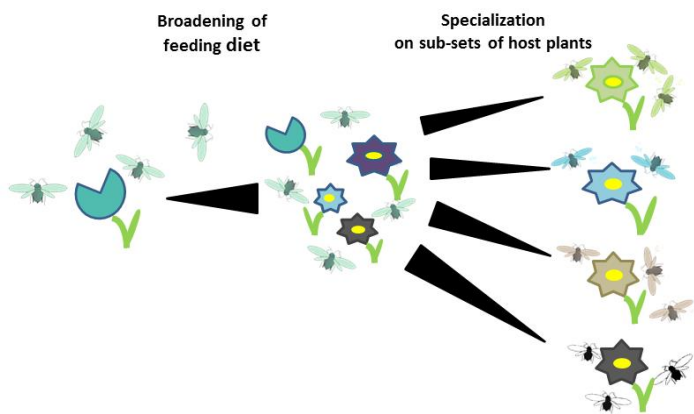

c) *Musical chairs* (Hardy & Otto 2014)

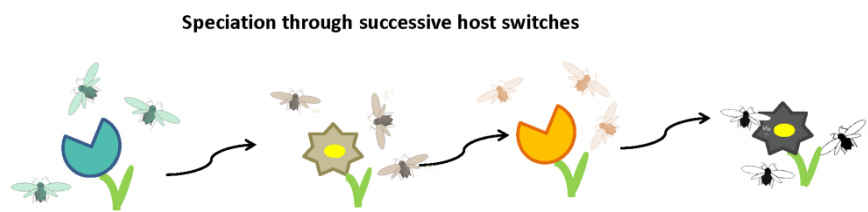

Figure 1: Schematic illustrations of the macroevolutionary scenarios of phytophagous insect diversification.

Population fragmentation and their specialisation onto newly captured host-plants then lead to the formation of new species. Hence this scenario explicitly predicts that species formation results from insect populations evolving towards the utilization of a restricted set of host-plants. Nevertheless, it suggests that this specialization process is often a consequence of the geographic isolation of generalist insect populations in areas inhabited by different host-plant species (Janz *et al.* 2006). Therefore the *Oscillation* hypothesis does not necessarily postulate that natural selection is the main driving force of species formation. However, subsequent papers quoting this scenario emphasize the central role of host-plant specialisation (Hardy & Otto 2014; Nakadai 2017; Wang *et al.* 2017).

Most of the predictions from the *Escape and Radiate* scenario are valid under the *Oscillation* hypothesis. However, the latter yields several new predictions (Table 1):

- **generalist*** diets should be "transient and repeatedly disappear in favour of specialization onto a limited set of related plants" (Nylin *et al.* 2014);

- the gains of new host plants are associated with host breadth expansion (Janz *et al.* 2001);

- the amplitude of the oscillation determines the number of potential host-plant species, therefore, insect clades with the most diverse host-use (the highest number of host-plant species) are expected to be more speciose than clades using fewer host species (Janz *et al.* 2006) and insect clades that encompass species that exhibit large host breadths should have higher speciation rates (Hardy & Otto 2014; Weingartner *et al.* 2006);

- shifts from a generalist diet to a specialist one should be associated with an acceleration of speciation rates; in other words, patterns of diversification should follow a model where cladogenetic events are associated with host breadth reduction (Hardy & Otto 2014);

- generalists have larger geographic ranges than specialists as they are able to colonize more habitats and can expand more easily (Slove & Janz 2011).

### I.3) The *Musical Chairs* (Figure 1c)

As opposed to the *Oscillation* hypothesis, Hardy and Otto (2014) proposed an alternative scenario in which speciation in herbivorous insects is driven by host-switching rather than transitions in diet breadth. The authors named their scenario *"**the Musical Chairs**"* (Hardy & Otto 2014). In this scenario, phytophagous insects speciate by the sequential capture of new host-plants and therefore speciation in a plant-feeding insect lineage is merely driven by the number of plants it can colonize. Host-driven speciation remains at the centre of this hypothesis.

The *Musical chairs* hypothesis yields several predictions that differentiate it from the previous macroevolutionary scenarios (Table 1):

- contrary to the *Escape* and *Radiate* hypothesis, the capture of a new host plant lineage does not initiate a radiation;

- contrary to the *Oscillation* hypothesis, gains of new hosts are not associated with host breath expansion (Hardy 2017), host breadth contraction is not associated with cladogenetic events (Hardy & Otto 2014) and overall there should be a negative correlation between host-plant breadth and speciation rates (Hardy & Otto 2014);

Table 1: Predictions from host-plant driven speciation scenarios: the first column indicates the evolutionary hypotheses tested and the headers of the other columns indicate phylogenetic comparative approaches used for testing them.

| Macroevolutionary scenarios | Insect and host tree comparison | Evolution of host associations | Evolution of host breadth | Insect diversification dynamics |
|---|---|---|---|---|
| *Speciation driven by host-plant specialization* | Cospeciation if insects play a role in their host-plant reproductive isolation | No overlap in host use among sister-species (Nyman *et al.* 2010) | Predominance of specialists over generalists (Janz *et al.* 2001, Winkler & Mitter 2008) | - herbivorous insects are more diverse than their non-herbivorous relatives (Mitter *et al.* 1988)<br>- the diversification dynamic of phytophagous insects follows a pattern of adaptive radiation (Janz 2011)<br>- the number of species within an insect clade positively correlates with the number of host-plant species (Janz 2006) |
| *Escape and radiate* | Phylogenetic tracking (Jermy 1976, Mitter & Brooks 1983) | Phylogenetic conservatism of host-plant lineages or host-plants with similar defences (Winkler & Mitter 2008) | No prediction | Increase in speciation rates upon the capture of new plant lineages or detoxification mechanisms (Wheat *et al.* 2007; Fordyce 2010) |
| *Oscillation* | Potentially phylogenetic tracking following the capture of a new host-plant lineage | Conservatism of host-plants following the capture of a new plant lineage. | - lability in host breadth (Janz *et al.* 2001, Janz & Nylin 2008)<br>- gains of new host lineages are preceded by host breadth expansion (Janz *et al.* 2001)<br>- positive correlation between diet breadth and geographic range (Slove & Janz 2011)<br>- speciation events associated with shifts from generalist to specialist (Hardy & Otto 2014) | - clades including generalist species are more speciose than clades with only specialists (Weingartner *et al.* 2006)<br>- speciation rates higher in lineages with labile host breadth (Hardy & Otto 2014) |
| *Musical chairs* | No prediction | Less conservatism of host plants in more speciose lineages (Hardy & Otto 2014) | Host breadth is not labile (Hardy & Otto 2014; Hardy 2017) | - negative association between speciation rates and host breadth (Hardy & Otto 2014)<br>- speciation rates positively correlate with host switching rates (Hardy & Otto 2014) |

- speciation rates are higher in insect lineages that exhibit **lability*** in host-plant associations.

The lack of connection between host breath contraction and speciation events and the negative association between host-breadth and speciation rates clearly differentiate the *Musical chairs* from the *Oscillation* hypothesis. However, several authors have pointed out that if generalist feeding diets are indeed ephemeral as expected when specialization towards host-plants is adaptive, it will be difficult to reconstruct its history accurately on phylogenetic trees (Hardy *et al.* 2016; Janz *et al.* 2016). Consequently, the relationships between host breath and speciation rates will be difficult to explore and the set of predictions that differentiate the *Musical chairs* from the *Oscillation* hypothesis will not always be testable.

Phylogenetic comparative methods (Pennell & Harmon 2013) including increasingly sophisticated diversification models (Beaulieu & O'Meara 2016; O'Meara & Beaulieu 2016; Rabosky 2006; Rabosky & Goldberg 2015; Stadler 2013; Stadler & Bokma 2013) can now be deployed to reconstruct ancestral character states, investigate the diversification dynamics of insect lineages and test whether shifts in diversification rates are associated with transitions in character states. Below we review how these methods have been used to investigate the evolution of plant / insect associations and test the predictions of host-driven speciation scenarios.

**II Phylogenetic approaches for testing ecological speciation scenarios**

**II.1) Comparing the phylogenies of plants and insects**

Several phylogenetic studies have compared the phylogenies of herbivorous insect and their host-plants. Some have investigated cospeciation patterns using dedicated tools such as tree reconciliation analyses (Conow *et al.* 2010; Page 1994) and distance-based methods for tree comparisons (Legendre *et al.* 2002). These tools statistically test the null hypothesis that the phylogenies of insects and their associated plants are more congruent than expected by chance and that speciation events are simultaneous. As stated above, this is only expected in very specific plant /insect interactions. As predicted by early taxonomic studies (Ramirez 1974; Wiebes 1979), cospeciation has been evidenced in figs and their phytophagous pollinating wasps but the degree of phylogenetic congruency observed varies according to taxonomic groups (see review by de Vienne *et al.* 2013 on cospeciation studies and Cruaud *et al.* 2012). Reciprocal adaptations of plants and insects (*i.e,* **coevolution***) have been unravelled in this study system (Jousselin *et al.* 2003; Weiblen 2004). However, it is not known whether the cospeciation patterns observed are the sole result of this coevolution, i.e. whether reciprocal selection exerted by both partners has driven the reproductive isolation of interacting populations (Althoff *et al.* 2014; Hembry *et al.* 2014), or whether matching speciation events have arisen through co-vicariance. In the other iconic model system for the study of plant/insect **coevolutionary diversification***, the *Yucca*–Yucca moth interaction (Pellmyr 2003), moth phylogenies parallel parts of the evolutionary history of their hosts. Some studies suggest that this pattern is the result of coevolution (Godsoe *et al.* 2009) while others hint towards co-vicariance (Althoff *et al.* 2012). In both *Yucca*-moths and fig-wasps the fact that the phytophagous insects are specific pollinators of their host-plants and breed inside them necessarily links the reproductive success of the two partners and increases the likelihood of concomitant speciation events. Therefore host-plant adapted traits are certainly pivotal in the speciation process of these insects.

Studies that have investigated **phylogenetic tracking*** between phytophagous insects and their host-plants have shown that insects rarely mimic their host-plant phylogenies (see reviews by de Vienne *et al.* 2013; Winkler & Mitter 2008 and Hsu et al. 2018 for a more recent study); a third of the 20 studies reported in these studies found partial patterns of congruence between insects and host plant phylogenies. In many plant/insect interactions, a simple observation of the patterns of distribution of host plants in the insect phylogeny can actually rule out cospeciation or phylogenetic tracking and studies on plant-insect interactions have moved away from cospeciation studies.

In order to test for successive burst of diversification in plants and insects, many authors have thus simply compared the timing of divergence of plants and associated insects (*e. g.*, Brandle et al. 2005; Gómez-Zurita *et al.* 2007; Kergoat *et al.* 2011; Kergoat *et al.* 2015; Leppanen *et al.* 2012; Lopez-Vaamonde *et al.* 2006; McKenna *et al.* 2009; McLeish *et al.* 2013; Pena & Wahlberg 2008; Segar *et al.* 2012; Stone *et al.* 2009; Vea & Grimaldi 2016; Wahlberg *et al.* 2013; Winter *et al.* 2017). Most of these studies suggest delayed colonization of already diversified groups of plants by insects groups at different temporal scales. They are generally based on mere qualitative comparisons of dates of divergence obtained from fossil calibrated phylogenies of both plants and associated insects, but can also include thorough statistical comparisons of dates obtained through phylogenetic methods (Loss-Oliveira *et al.* 2012; McLeish *et al.* 2013). They are often framed as supporting the *Escape and Radiate* theory. However these studies do not give any information on the speciation process behind the diversification of the insect lineages studied, they merely indicate the timing of host plant colonization. The comparison of the **diversification dynamics**[*] of both herbivorous insects and their host-plants provide a more direct test of host-driven speciation hypotheses: under host-driven adaptive radiation, insect diversification dynamic is expected to roughly follow the diversification of its host-plant lineage. This can be investigated through diversification analyses (see II. 3.2) but can also include the comparison of the fossil records of both insects and plants (Labandeira & Currano 2013; Labandeira *et al.* 1994). In general, the studies of fossil assemblage are decoupled from phylogenetic studies of plant/insect associations. We advocate for combining fossil and phylogenetic evidence whenever possible.

**II. 2) Reconstructing the evolutionary trajectory of traits involved in host-plant use**

II.2.1) Evolution of host associations

Phylogenetic inferences are widely used to reconstruct the evolutionary trajectories of phenotypic traits throughout the diversification of a lineage. Most phylogenetic studies of phytophagous insects map the history of host association onto the resulting trees –at different taxonomic levels (host plant order, family, genus depending on the level of host specialization of the insect clade studied and multistate characters are used when species are polyphagous).

These reconstructions generally suggest a pattern of **host conservatism**\*. This assertion stems from mere observations of the reconstructions but numerous studies conduct statistical tests. These include the permutation tail probability test (PTP, Faith & Cranston 1991) or some index of phylogenetic signal such as ʎ's of Pagel (1999) (see Winkler & Mitter 2008 for a review on host plant conservatism patterns and Leppanen *et al.* 2012; Wilson *et al.* 2012, Hernández-Vera *et al.* 2019 for more recent examples).

Host conservatism is often interpreted as following the predictions of *Escape and Radiate* and therefore evidence that speciation was promoted by host-plant specialization. However, showing that related insects feed on related plants does not say much about the process that has generated this pattern nor connects mechanistically host-plant use evolution to speciation. The use of vague wording such as host-associations *"favour"* or *"constrain"* speciation is commonly found when discussing host-conservatism in the literature and it is difficult to conclude from these studies that specialization towards one or a few plant species is the main pathway towards the formation of new phytophagous insect species. The pattern of host conservatism is in agreement with a scenario in which insects have radiated onto a plant lineage but it could also suggest that host-plant shifts are not important promoters of speciation events.

A more direct estimation of the contribution of host-plant adaptation to the speciation process consists in inferring the frequency of host-plant shifts in relation to speciation events. If adaptation to different ranges of host-plants drives reproductive isolation and speciation, it follows that insect **sister species**\* should partition host-plant resources: i.e. they should show no or little overlap in the plant species they use.

To test this hypothesis, early studies have conducted sister species comparisons of host ranges and more recent studies have reconstructed the evolution of insect ecological niches (defined as the combination of feeding habits and host-plant species) and estimated the number of niche shifts associated with speciation events (Table 2.1). Comparisons of alternative models of evolution of host-use have also been conducted. All studies but one showed that the numbers of niche shifts observed generally represented less than 50% of the speciation events and were lower than expected if the niches were randomized onto the phylogeny. This suggests that ecological speciation is not the main process behind the diversification of these lineages. Authors have also observed that host-use differentiation occurred at the root of the trees and therefore concluded that it played a minor role in recent speciation events (Table 2.1).

We must keep in mind that these methods may overlook many host shifts (shifts that resulted in population extinction). Therefore, the studies that estimate niche differentiation at speciation events probably present an overestimation of the impact of host shifts in speciation. It is nevertheless surprising that such studies have not been conducted on more study systems. This is likely due to the fact that they require a precise knowledge of the range of host-plants used by each insect species.

Table 2: Summary of studies testing the predictions of host-driven speciation scenarios using phylogenetic methods. Cospeciation studies and studies on host-plant conservatism are not included as they are less recent and already synthetized in respectively: de Vienne *et al.* 2013 and Winkler & Mitter 2008.

| Predictions tested | Taxa | Reference | Approaches | Results | Conclusions |
|---|---|---|---|---|---|
| 1) Partitioning of host plants at speciation events | *Aphanartum* (25 spp.)  | Jordal & M Hewitt 2004 | Sister species comparison of host ranges | ∅ | Geographic isolation is more important than host switching in speciation events |
| | Nematinae (125 spp.)  | Nyman *et al.* 2010 | Investigate niche shifts at speciation nodes through ancestral state reconstruction | ∅ | |
| | *Blepharoneura* (49 spp.)  | Winkler *et al.* 2018 | | ∅ | |
| | *Cinara* (76 spp.)  | Jousselin *et al.* 2013 | | ∅ | |
| | *Caloptilia* (13 spp.)  | Nakadai & Kawakita 2016 | Niche dissimilarity index through time | ∅ | |
| | *Neodiprion* (19 spp.)  | Linnen & Farrell 2010 | Comparisons of models of evolution of host use (speciational model vs gradual) | √ | Host shift (in allopatry) induces speciation |
| 2) Host breadth is labile | *Dendroctonus* (19 spp.)  | Kelley & Farrell 1998 | Phylogenetic signal of host breadth | √ | Repeated broadening of host breadth |
| | Lymantrinae (55 genera)  | Wang *et al.* 2017 | Phylogenetic signal of host breadth and transition rates estimation | √ | **Oscillation** |
| | Nymphalidae (551 spp.)  | Nylin *et al.* 2014 | Phylogenetic signal of host breadth and transition rates estimation | √ | |
| | Nymphalini (31 spp.)  | Janz *et al.* 2001 | Visualization of reconstruction of host breadth | √ | |
| | 15 insect groups | Nosil & Mooers 2005 | Estimation of transition rates in host breadth | √ | Specialization is not a dead-end |
| | *Boloria* (37 spp.)  | Simonsen *et al.* 2010 | Visualization of reconstruction of host breadth | √ | Repeated broadening of host breadth, **oscillation** |
| | Nymphalidae  | Hamm & Fordyce 2016 | Phylogenetic signal of host breadth | ∅ | **No oscillation** |
| 3) Changes in host breadth spur diversification or speciation events are associated with shift away from polyphagy | Papilionoidea (2573 spp.)  | Hardy & Otto 2014 | Use of BiSSEness to test whether change in host breadth this associated with speciation events (speciational model vs gradual model of evolution for host breadth) | ∅ | **Musical chairs** |
| | Nymphalini (31 spp.)  | Janz *et al.* 2001 | Estimate whether polyphagy is ancestral, estimate gains of new host vs losses of hosts (if polyphagy drives speciation; gains should exceed losses) | √ | **Oscillation** |
| | Nymphalini (172 spp)  | Hardy 2017 | Use of DEC model to reconstruct the history of host use and estimate gains and losses of plants | ∅ | Speciation not associated with shift away from polyphagy, **no oscillation** |
| 4) Detoxification mechanisms in insects evolve in response to plant association | Blepharidae (23 spp.)  | Becerra 1997 | Comparison of insect phylogeny with chemical defense similarity dendrogram of host plants | √ | Coevolutionary arm race |
| | 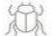 Blepharidae (37 spp.) | Becerra 2003 | Comparison of timing of acquisition of plant defenses and insect counter-defenses | √ | Coevolutionary arm race |
| | 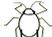 Lygaeinae (20 spp.) | Bramer *et al.* 2015 | Reconstruction of the ability to resist or sequester cardenolides | √ | Adaptation of insects to host plant defenses |
| | 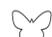 *Heliconius* | de Castro *et al.* 2018 | Review on correlated evolution of anti-herbivory adaptations in plants and counter-adaptations in *Heliconius* | √ | Adaptation of insects to host plant defenses |
| | 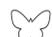 Melitaeini (77 spp.) | Wahlberg 2001 | Reconstruction of host association on insects and the presence of glycosides in associated plants | √ | Insects switch to chemically similar plants |
| | 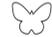 174 taxa | Endara *et al.* 2017 | Comparison of Lepidoptera assemblages associated with different plant species | √ | Similarity of assemblages on chemically similar plants; host association driven by similarity of plants |
| | 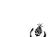 Apocynaceae | Livshultz *et al.* 2018 | Evolution of cardenolide production in plants | √ | Plant defenses evolve in response to herbivorous insect predation |
| 5) Diversification rates in phytophagous insect clades > non phytophagous insects | 13 families in various orders | Mitter *et al.* 1988 | Sister groups comparison of phytophagous vs non phytophagous clades | √ | Phytophagy promotes diversification |
| | 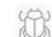 1900 spp | Hunt *et al.* 2007 | Sister groups comparison of phytophagous vs non phytophagousclades and estimation of diversification rates | Ø | Various types of niche shifts explain beetles diversification |
| | 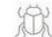 Erolytidae, 53 taxa | Leschen & Buckley 2007 | Correlated evolution between species richness and phytophagy | Ø | |
| | 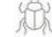 367 spp, 172 families | McKenna *et al.* 2015 | Infer shifts in diversification rates (MEDUSA) | √ & Ø | Some shifts associated with phytophagy others not |
| | 31 insect orders | Wiens *et al.* 2015 | Phylogenetic regression | √ & Ø | Different ecological factors prevail at different scales: phytophagy promotes diversification overall but not in all orders |
| 6) Major host shifts spur diversification | 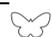 Nymphalidae (54 genera) | Nylin & Wahlberg 2008 | Estimation of diversification rates after two major host shifts | √ | **Escape and radiate and/or Oscillation** |
| | 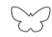 Butterflies (15 groups) | Fordyce 2010 | Test for shifts in diversification rates across the phylogeny (LASER) | √ | Burst of diversification concomitant to some host shift, **Escape and radiate** |
| | 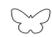 *Adelpha* (38 spp.) | Mullen *et al.* 2011 | Test for shifts in diversification rates across the phylogeny (SymmeTREE) | √ | One rate shift attributed to hots shift but also other ecological factors |
| | 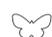 *Adelpha* & *Limenitis* (200 spp.) | Ebel *et al.* 2015 | Test for shifts in diversification rates across the phylogeny (BAMM) | √ | Shift to Rubiacea played a role in insect diversification |
| | 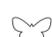 Nymphalidae (398 genera) | Pena & Espeland 2015 | Detection of shifts in diversification rates (MEDUSA & BiSSE with trait= feeding on a particular plant family) | √ & Ø | Shift to Solanacea spurred diversification |
| | Hesperidae (290 genera) | Sahoo *et al.* 2017 | Detection of shifts in diversification rates LASER, BAMM, BiSSE (trait= feeding on monocotyledonss vs dicotyledons), HiSSE | √ | Several diversification shifts however could be associated with grassland expansion and geographic factors |
| | Symphita (164 spp.) | Nyman *et al.* 2019 | Detection of shifts in diversification rates MEDUSA, BAMM, MuSSE (traits =feeding on angiosperms, gymnosperms, pteridophytes) | √ & Ø | No significant shifts associated with feeding on angiosperm, discuss confounding factors |
| 7) Acquisition of detoxification mechanisms spur diversification in insects, new defenses spur diversification in plants | Pierinae (60 spp.) | Wheat *et al.* 2007 | Distribution of a defense mechanism in insect and assessment of the glucosinolate in their host / comparison of rates of diversification in two sister clades with and without this defense | √ | **Escape and Radiate** |
| | Pieridae (96 spp.)/Brassicales | Edger *et al.* 2015 | Distribution of of a defense mechanism and estimation of shifts in diversification rates in plants and insects (MEDUSA) | √ | Complex coevolutionary arm race involving gene duplication in plants and associated with shift in diversification |
| | Asclepias | Agrawal *et al.* 2009 | Comparison of various model of evolution for plant defences including a speciational model | √ | Plant defenses are costly and can hinder diversification |
| 8) Adaptive radiation patterns | *Erebia* (74 spp.) | Pena *et al.* 2015 | Test for shifts in diversification rates (BAMM & DDD) | √ & Ø | Density dependant model fits the data, diversification shifts not always significant depending on methods : acceleration of diversification associated to colonization of new area |
| | *Cinara* (92 spp.) | Meseguer *et al.* 2015 | Test for shifts in diversification rates (TreePar) | Ø | Constant rate of diversification, no adaptive radiation |
| | *Blepharoneura* (49 spp.) | Winkler *et al.* 2018 | Test for shifts in diversification (LASER, DDD) | Ø | Constant rate of diversification, no adaptive radiation, diversification patterns mostly explained by geographical factors |
| | Sesamiina (241 spp.) | Kergoat *et al.* 2018 | Test for shifts in diversification through different methods (BAMM, *SSE) | Ø | Inverse patterns of diversification in insects and associated plants ; host plant diversity alone does not explain insect diversification |
| 9) Positive relationship between insect lineage diversity and their host plant diversity | 115 spp. | Farrell & Mitter 1998 | Sister-clade comparison of angiosperm vs non angiosperm feeding groups | √ | Host-driven speciation |
| | Nymphalidae (309 genera) | Janz *et al.* 2006 | Sister clade comparisons | √ | **Oscillation hypothesis** |
| | Lymantrinae (55 genera) | Wang *et al.* 2017 | Phylogenetic regression | √ | |
| | 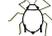 Coccidae (158 genera) | Lin *et al.* 2015 | ⎫ Regression without phylogenetic correction | √ | ⎫ Host-driven speciation |
| | 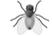 Cecidomyiid (352 genera) | Joy & Crespi 2012 | ⎭ | √ | ⎭ |
| 10) Clades including generalists speciate faster than clades with only specialists; diversification rates are positively correlated with host breadth | 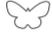 *Nymphalis* & *Polygonia* (20 spp.) | Weingartner *et al.* 2006 | Sister clade comparison of species richness between clades with only specialist and clades that encompass species with large host breadth | √ | Broadening of host range: plasticity facilitates the capture of new hosts and subsequent ecological speciation |
| | 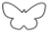 Nymphalidae (54 genera) | Nylin & Wahlberg 2008 | Estimation of diversification rates | √ | Clades experiencing higher rates of diversification experience a polyphagous state: **oscillation** |
| | 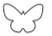 Papilionoidea (2573 spp.) | Hardy & Otto 2014 | ⎫ | Ø | **No oscillation, Musical chairs** |
| | 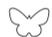 Nymphalidae | Hamm & Fordyce 2015 Hamm & Fordyce 2016 | Test for shifts in diversification through different *SSE methods (trait= host breadth) | Ø | Host breadth dynamics does not drive diversification |
| | 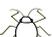 Coccidea | Hardy *et al.* 2016 | ⎭ | √ | Specialization by drift |
| | 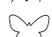 Nymphalini (172 spp.) | Hardy 2017 | Phylogenetic independant contrasts (proportion of generalists vs number of species in a genus) | Ø | **No oscillation** |
| | 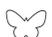 Sesamiina (241 spp.) | Kergoat *et al.* 2018 | Evaluate whether shifts in diversification rates follow shifts in host breadth (BAMM and others) | Ø | **Oscillation** might not be detectable at this scale |
| 11) Diversification rates of clades with labile host association> diversification rates of clades with conservatism in host association | 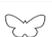 Papilionoidea (2573 spp.) | Hardy & Otto 2014 | Use of *SSE to test whether speciation rates vary between lineages, correlation between host switching rates and diversification rates | √ | **Musical chairs** |

Coleoptera,  Lepidoptera,  Hemiptera,  Diptera,  Hymenoptera

√ Prediction verified, Ø prediction not verified

DEC (Ree, Smith 2008); Laser (Rabosky 2006); Medusa (Alfaro et al. 2009), BiSSE (Maddison, Midford, Otto 2007), BAMM (Rabosky et al. 2014)-, DDD (Etienne et al. 2012), HiSSE (Beaulieu, O'Meara 2016), SymmeTREE (Chan, Moore 2002)

II.2.2) Host-breadth evolution

In order to investigate the role of specialization in insect species formation, phylogenetic studies have also investigated the distribution of host breadth throughout the evolutionary history of insects groups. According to the *Oscillation* hypothesis, host breadth should vary along the phylogeny of insects and the character state "generalist" should be transient; i.e. phylogenetic reconstructions should show many transitions between specialists and generalists. Many studies have indeed demonstrated the lability of host breadths (Table 2.2), which is compatible with the *Oscillation* hypothesis but does not prove it.

In order to test whether these changes are linked with speciation, some studies have investigated how gains and losses of new hosts along the phylogenetic history of lineage relate to the number of speciation events (Table 2.3), they obtained mixed support for alternative scenarios.

Maybe another way to investigate the link between 'host-breadth changes' and 'gains of new hosts' that underlies the *Oscillation* hypothesis, without using diversification analyses (see II.3.3) would be to test for the correlated evolution of these two characters; i.e. explore how often the gain of new hosts is associated with a transition from specialist to generalist.

II.2.3) Evolution of traits involved in host-plant choice

Several studies that aimed at finding support for the coevolutionary arm race hypothesized by the *Escape and Radiate* scenario have investigated the evolution of detoxification mechanisms in insect lineages and have shown that they correlate with changes in their host plant defences (Table 2.4). Studies that show that host switches occur between plants with similar defenses rather than between closely related plants generally conclude that a coevolutionary arm race occur between insects and plants. However some of these studies argue in favour of a scenario in which herbivores simply "choose" host-plants based on their own defensive traits. Host switches induce insect species differentiation but these are not the product of coevolutionary interactions.

Reconstructions of the history of plant defences have also been conducted on plant phylogenies. However direct tests of acceleration of diversification upon the acquisition of new defence mechanisms sometimes contradict the *Escape and Radiate* prediction and suggests that investment in costly defences can impede plant diversification (Table 2.4).

In contrast to detoxification mechanisms, traits involved in host recognition and host-plant choice (chemosensory traits) have been less studied in a phylogenetic context. However their evolution probably plays as important a role as adaptations to plant defences in phytophagous insect speciation (Smadja & Butlin 2009). Matsuo (2008) showed that an odour binding protein can evolve relatively fast in closely related *Drosophila* species through gene duplications and losses, and proposed that this dynamic could explain the evolution of host preferences in this species complex. Comparative analyses of odour binding proteins and chemosensory proteins from the genomes of several Arthropoda species (Sánchez-Gracia *et al.* 2009; Vieira & Rozas 2011) revealed a high number of gains and losses of genes, pseudogenes, and independent origins of gene subfamilies. This dynamic, if analysed in relation to host choices and host breadth in a phylogenetic context, could explain some host shifts and subsequent speciation events. Finally, deciphering the evolutionary dynamics of genes involved in mate recognition and their link with host association could also inform us on the role of host-plants in the speciation of insects that feed (and often mate) on them. For instance, Schwander *et al.* (2013) showed that cuticular hydrocarbon profiles involved in mate choices vary among *Timema* species (Phasmatodea), and that most evolutionary changes in hydrocarbon profiles occur in association with host-plant shifts and speciation events in this genus of phytophagous insects.

In summary, many studies have investigated the evolution of detoxification mechanisms in insects, framing their hypotheses within the *Escape and Radiate* theory. Studies investigating changes in traits involved in host recognition throughout the diversification of insects are rare and they are generally not framed within host-plant driven scenarios. Such studies are needed in order to determine which traits underlie host-plant associations and whether their evolution drives speciation events (see Perspectives).

**II. 3) Studying how host-plant uses impact the diversification dynamics of herbivorous insects**

Methods for testing diversification dynamics have expanded over the last decade. Testing for the existence of temporal bursts of diversification was once restricted to analyses of groups with comprehensive fossil records. Diversification dynamics can now be studied through comprehensive phylogenies (Goswami *et al.* 2016). Given the breadth of available methods, theoretically, nearly all hypotheses can be put to test.

<u>II.3.1) Are phytophagous insects more diversified than their related counterparts?</u>

Studies that posit that host-plant adaptation favours phytophagous insect diversification predict that those are much more diversified than non-phytophagous insects. However this assertion deserves to be statistically tested. The first study addressing this question suggested that herbivorous clades contain more species than their non-phytophagous sister clades (Table 2.5). However, comprehensive phylogenies associated with recent comparative methods suggest that several types of niche shifts beside phytophagy can explain insect species richness. Studies within insect orders have mostly been addressed within Coleoptera –as they are by far the most diverse order, but more in-depth analyses of other orders would also be insightful.

To our knowledge likelihood-based character state dependent diversification models (known as, the *SSE models, such as BiSSE, ClaSSE and QuaSSE for binary, multistate and quantitative traits, respectively FitzJohn 2010; Maddison & FitzJohn 2015; Maddison *et al.* 2007), which test whether transition in character states are associated with variations in speciation and extinction rates, have not been yet used to test the role of phytophagy in insect diversification. A robust estimate of diversification parameters by these methods requires dense and random species sampling throughout the phylogeny (more than 15% of the species must be included in the phylogeny in order to conduct unbiased tests, FitzJohn 2010). Therefore *SSE tests await the availability of comprehensive phylogenies in more insect orders. Nevertheless, a caveat of these diversification models is that they may overlook more complex models involving many unmeasured and co-distributed traits. In particular, for phytophagy, other traits that can drive shifts in diversification may cause a spurious detection of increased diversification rates in phytophagous insects if these trait's states partly correlate with phytophagy, or may instead erase any signal due to phytophagy. The HiSSE model (Beaulieu & O'Meara 2016) which models hidden characters that influence diversification might help untangling these confounding factors.

### II.3.2) Testing for adaptive radiation

Diversification analyses can also be applied to test whether the capture of new host-plants has favoured adaptive radiations. Under such a scenario, the diversification curve of phytophagous insect clades should exhibit early bursts of speciation upon the capture of new groups of host-plants. Insect lineages are eventually expected to fill the newly found niche space and the diversification curves should reach a plateau.

Several studies have investigated acceleration in speciation rates in insect lineages upon the capture of new host-plant lineages (Table 2.6) or detoxification mechanism (Table 2.7). These studies have mainly been conducted in butterflies. Studies have rarely included trait-dependent diversification models and when they have, they have often relied on a few colonization events that are correlated with important geographical changes, making it difficult to conclude on the causal effect of the sole host shifts (Table 2.6).

Among studies testing adaptive radiation patterns (Table 2.8), few have investigated whether the number of species reaches a plateau after an initial burst of speciation. This is better addressed using density-dependent models of diversification (DDD) that test whether rates of speciation decrease once the number of species supposedly reach the carrying capacity of the novel ecological niche (Etienne *et al.* 2012; Rabosky & Lovette 2008). But these tests can only be applied to lineages in which the number of species in each subclade is well known. In any case, studies generally provided weak support for adaptive radiation scenarios (Table 2.8). They all suggested that the diversification patterns of phytophagous insects cannot solely be explained by the availability of suitable host-plants and suggested that abiotic factors such as geography and temperature should be taken into account in diversification scenarios.

### II.3.3) Correlating host breadth with diversification dynamic

Advocates of the *Oscillation* hypothesis suggest that clades showing a higher diversity of host-use (using more host-plant species altogether) should be more diverse than their sister clade (Janz *et al.* 2006), and this prediction has been verified in butterflies, Coccidae and some gall inducing flies (Table 2.9). Although these results support a model where the diversity of phytophagous insects is sustained by the diversity of the hosts they use; they could fit both a model of *Oscillation* and the *Musical chairs* hypothesis. To tell apart the *Oscillation* from the *Musical chairs*, it is more informative to test how host breadth variations affect insect diversification dynamics.

According to the *Oscillation* hypothesis, clades including generalist species should be more speciose than clades including only specialists. Early papers have used fairly basic methods such as 'sister clade analyses" to test this prediction while more recent investigations have adopted trait-dependant diversification models (Table 2.10). These studies provided mixed support for *Oscillation* scenarios. Furthermore, even when a specific prediction was met, the authors could not always reach a firm conclusion: as previously mentioned, the transient nature of the generalist feeding diet under host-driven speciation scenarios complicates predictions. The negative or positive relationship observed between host breadth dynamics and diversification can always be obscured by the rapid evolution of host breadth (Janz *et al.* 2016; Nylin & Janz 2009). The way host breadth is measured (binary vs continuous) is also known to affect the results (Hamm & Fordyce 2015) and it has been long recognized that categorizing species into either specialist or generalist can be somewhat subjective (Janz *et al.* 2001). Furthermore, although the *Musical chairs* yield specific predictions, those are mainly rebuttals of the *Oscillation* predictions (but see Table 2.11). Yet, rejecting an *Oscillation* scenario does not instantly mean that a *Musical chairs* scenario is at play. Hence despite the existence of sophisticated methods and comprehensive phylogenies, the prevalence of these scenarios in insects is still debated.

Another caveat of these studies lies in the distinction between speciation and extinction rates. Indeed, if specialization towards host plants can accelerate speciation it can also increase extinction risks when host plants are not highly abundant. Consequently, predicting exactly how changes in host-plant breadth affect diversification rates is difficult. Finally diversification methods such as *SSE models are known to generate type I errors (Bouchenak-Khelladi *et al.* 2015; Davis *et al.* 2013; Rabosky & Goldberg 2015), and can thus inflate the role of host plants and host breadth in diversification when those are investigated. It is therefore, highly recommended to conduct the analyses on a set of randomized trees in order to test whether the constrained diversification model (*i.e.* the model in which evolutionary transitions in character states are associated with shifts in extinction and/or speciation rates) is also chosen in these analyses (as in Hardy *et al.* 2016).

Moving away from methodological issues, our survey also underlines that many studies that explore macroevolutionary scenarios have been conducted on Lepidoptera. In order to have a better understanding of the role of host-plant shifts in insect diversification, it seems necessary to test the predictions of macroevolutionary scenarios on more insect groups. Aphids (Hemiptera) could be good candidates for such investigations. Their range of host-plants is well documented (Blackman & Eastop 2006; Holman 2009). Although most aphids are host-specific, some species are polyphagous. Some species are even only polyphagous during part of their life-cycle. This temporary broadening of diet has already been suggested to facilitate the capture of new host-plants (Moran 1992) and subsequent speciation onto these new lineages. These life-cycle transitions could have thus favoured aphid diversification (Jousselin *et al.* 2010; Moran 1992). This scenario actually fits the *Oscillation* hypothesis and should be explored. Other Hemiptera which host plant repertoire is well known such as Coccidae (Garcia-Morales et al. 2016) and psyllids (Ouvrard et al. 2015) could also be used to test the predictions on the role of host breadth evolution in macroevolutionary scenarios. The limitation for these groups for now lies in the availability of robust and comprehensive phylogenies.

### III Perspectives

As seen throughout this review, phylogenetic comparative methods provide the template to test hypotheses on the role of host plant association in the speciation of phytophagous insects. While those methods have undoubtedly advanced the field since the *Escape and Radiate* paper, one must keep in mind that phylogenetic comparative methods often constitute mere correlation analyses. Significant associations between character changes and cladogenetic events might arise as a consequence of speciation itself when post-speciational character changes occur. Furthermore, comparisons of models of evolution such as those used in trait-dependent diversification analyses often rely on trees that encompass few transitions in character states and are therefore not always robust (see Beaulieu & Donoghue 2013). In such analyses, the "best model" is not necessarily the true model and significant *P* values should not be interpreted as strong evidence for an evolutionary scenario.

In addition to using the approaches focused on host-plant associations and diet breadth cited throughout this review, one way to further investigate hypotheses of speciation driven by associations with host-plants would be to integrate a variety of data in a phylogenetic context. Below we outline three potential directions for future research: 1) disentangling the role of plant-insect interaction from that of co-variates, such as geography and climate; 2) studying traits and genes underlying the association; 3) combining phylogenetic analyses with interaction network approaches including other partners, at various ecological scales (from community-scale to global scale).

#### III.1) Investigating the role of abiotic factors: geography and climate

Geography and ecology are always closely intertwined in speciation scenarios. There have been several studies that have investigated geographic range expansion in herbivorous insects (Becerra & Venable 1999; Slove & Janz 2011); climate induced host shifts (see Nyman *et al.* 2012 for a review and more recently Lisa De-Silva, 2017; Owen *et al.* 2017; Pitteloud *et al.* 2017; Sahoo *et al.* 2017; Sanchez-Guillen *et al.* 2016; Winkler *et al.* 2009) and climate driven diversification dynamics (Kergoat *et al.* 2018). All these studies suggest that abiotic factors are entangled with host-plant changes in species diversification scenarios. However there are few studies that explicitly test the predictions of speciation through geographic isolation (Barraclough & Vogler 2000) and whether these events systematically accompany host shifts or sustain speciation events (but see Jordal & Hewitt 2004; Jousselin *et al.* 2013; Doorenweerd *et al.* 2015; Hardy *et al.* 2016). Such analyses are important if we want to tell whether adaptations to new host plants represent post-speciational changes following geographic isolation rather than the main driver of speciation events. Cospeciation methods that take into account the biogeographic history of interacting lineages (Berry et al. 2018) could also be used to investigate whether host shifts are associated with dispersal events in systems where host-plant and insect phylogenies show congruent patterns.

### III.2) Unravelling traits involved in the interaction and their underlying genes, and integrating this information in phylogenetic studies

Interactions between insects and their host-plant are ultimately mediated by traits, such as host-plant defences and the capacity of circumventing plants defences, but also host-plant cues and the capacity for herbivores to detect those cues. Characterizing such traits, their genetic determinism and looking at their evolutionary trajectory would greatly advance our understanding of the diversification of insects (*e. g.*, de Castro *et al.* 2018). Testing whether different trait states are associated with different speciation rates can be performed using *SSE methods (*e. g.*, as in Onstein *et al.* 2017 for a trait related to frugivory in palm trees). In addition, methods that test whether patterns of trait evolution conform to a model accounting for interactions mediated by those traits are currently being developed (Manceau *et al.* 2016; Drury *et al.* 2017). They could shed lights on the processes underlying herbivore diversification. However, targeting traits involved in plant-insect interactions may be challenging. Pivotal traits are difficult to identify; they include chemical, behavioral and metabolic traits and when they are properly characterized they are often multigenic.

Perhaps a promising direction for future research is the implementation of a hybrid genomic approach that combines transcriptomics, phylogenomics, comparative analyses and population genomics (see Nevado *et al.* 2016). In such approaches, full transcriptomes of species from a target clade (for instance, a clade of phytophagous insect) are generated. These transcriptomes are used to generate a phylogeny, where diversification and character evolution tests can be performed (*i.e.* test for diversity-dependent diversification, shifts in diversification following host-plant shift). Then, genes under selection can be detected from transcriptomic data using population genetics statistics, and can be matched to existing databases (e. g., Lepbase for Lepidoptera, Challis *et al.* 2016) for identification purpose. Additionally, genes that are down or upregulated can also be detected by classical tests of differential expression and identified, and the association of genes under selection, either via different sequence or expression pattern, with species diversity can be tested. One of the limits of this approach lies in the availability of specimens (transcriptomic data need to be obtained from fresh or suitably preserved tissues; biological replicates are needed): those might be difficult to obtain throughout an entire phylogenetic tree. The availability and quality of the reference gene database to match genes with putative functions might also limit the applications of this approach.

### III.3) Combining phylogenetic with interaction network approaches, at various ecological scales

Herbivores and the plants they feed on form interaction networks, and as such the structures of the networks can be characterized by several parameters, such as modularity (the propensity of a group of species to interact with a similar set of partners) and nestedness (the propensity of specialist species to interact with generalist species and vice-versa). Antagonistic interaction networks, such as plant-herbivore networks, tend to be highly modular (Thébault & Fontaine 2010). A recent study combining interaction network with phylogenetic approaches on simulated and real datasets predicted that the *Escape and*

*Radiate* scenario should produce a modular network structure, whereas the *Oscillation* scenario should produce a more nested structure (Braga *et al.* 2018). When applied to real data (two butterfly families, Nymphalidae and Pieridae), this approach revealed that host-plant butterfly networks tend to be both modular and nested, which the authors interpret as being the result of a complex pattern of diversification, involving both episodes of radiation on new hosts (producing modules containing closely related species) and occasional shifts to other host lineages, producing both nestedness within modules and connections between modules.

Additionally, phylogenetic and network approaches could be expanded to encompass other interacting partners (*e. g.,* Ives & Godfray 2006). Indeed, insect-host-plant communities can be seen as ecosystems where biotic interactions, such as parasitism and mutualism also take place (Forister et al. 2012). These partners can indirectly influence the interaction between plants and their herbivores: *e.g.*, direct competition (Jermy 1988) apparent competition between herbivores, stemming from shared natural enemies (Holt 1977), and vice-versa (*e. g.,* when herbivory elicits anti-herbivore defences mediated by herbivore enemies, Fatouros et al. 2008). Multitrophic interactions probably explain many diversification patterns in herbivorous insects (Singer, Stireman 2005).

Finally, such approaches could be applied both at a large scale (*e. g*., Braga et al. 2018), to embrace global patterns of diversification and interaction, or at the community level (Ives, Godfray 2006; Elias *et al.* 2013), where interactions actually occur, and where fine-scale processes (*e. g.*, host-plant shift at the species or the population level) can be unveiled.

**Conclusions**

Many phylogenetic studies of plant-insect associations now include formal tests of macroevolutionary scenarios involving host-driven speciation. In an attempt to summarize the literature on this topic, we show that the predictions of host-plant driven speciation are not straightforward and can vary depending on studies. We advocate a standardization of these predictions to facilitate cross study analyses. Furthermore, it is also recognized that different scenarios can leave the same phylogenetic signature (Janz *et al.* 2016) and that depending on the analytical approaches undertaken to test the predictions laid out in Table 1, conclusions can vary (Table 2). Unfortunately this means that the interpretations of phylogenetic inferences can remain somewhat subjective. But these shortcomings should not cloud the progresses that have been made in the field. Phylogenetic comparative analyses help framing hypotheses and clarify some of the narratives used to explain the diversification of phytophagous insects. In order to move towards a standardization of phylogenetic approaches, we propose here a list of relatively simple tests that could be applied to an insect phylogeny to test some of the non-controversial predictions of host-driven speciation scenarios (Fig. 2). The limitations of these approaches have been described throughout this review.

Finally, this survey of the literature shows that: 1) the simple assumption that phytophagy has accelerated insect diversification is not always sustained by meta-analyses; 2) the expectation that sister lineage will use different ranges of host plants is not often tested, and, when it is, the predictions of a host-driven speciation scenarios are not often met. We then underline that the results of phylogenetic comparative methods cannot be interpreted as hard evidence as they remain mere correlations. In the end, a full understanding of the processes explaining the diversification of phytophagous insects will require the integration of phylogenies with other data sources and analytical methods.

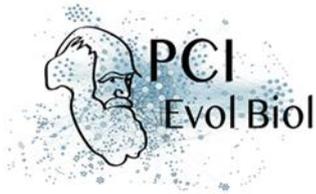

To conclude, if the last two decades have seen the rise of molecular phylogenies and the development of analytical methods that include ecological data, this should not mask the need for thorough curation of the data used before applying any phylogenetic comparative analyses. Qualifying host associations of insect species necessitates field work and advanced taxonomy, as mistakes can seriously impact the results of macroevolutionary studies. Functional studies aimed at deciphering host-plant adapted traits in insects (and in particular traits implied in host choice) and characterizing genes that underlie these traits are also needed to integrate this data in a phylogenetic context and link microevolutionary processes with macroevolutionary scenarios.

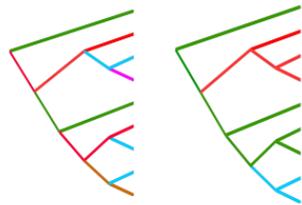 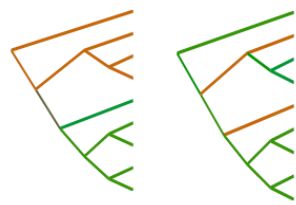 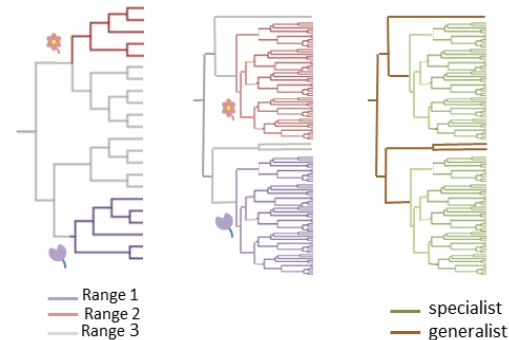

Figure 2: Suggestions of phylogenetic comparative methods that can be deployed to test some of the prediction of host-driven speciation scenarios: a) tests that rely on the reconstruction of host-plant range evolution; b) tests that rely on the reconstruction of host-breadth; c) diversification analyses that rely on reconstruction of both these characters. H1, H2, H3 are alternative scenarios and are represented by schematic phylogenetic reconstructions.

> **Glossary**
>
> **Adaptive radiation**: the evolution of ecological and phenotypic diversity within a rapidly multiplying lineage (Schluter 2000). It occurs when natural selection drives the fast divergence of an ancestral species into multiple descendants that exploit different ecological niches.
>
> **Coevolution:** reciprocal evolutionary changes occurring in two or more species that result from reciprocal selective pressures exerted by the interacting partners.
>
> **Coevolutionary diversification**: when diversification patterns arise from coevolution.
>
> **Cospeciation:** simultaneous speciation events in lineages involved in long-term interspecific associations which result in congruent phylogenies and temporal congruence of speciation events.
>
> **Diversification dynamic:** rates of species formation and extinction through time.
>
> **Ecological specialization**: when species are limited to a restricted set of resources (diet- habitat-niches), as a result of evolutionary trade-offs.
>
> **Evolutionary lability**: tendency for a character to change frequently throughout evolution.
>
> **Generalists:** species that use a wide niche (a wide range of host plants belonging to different lineages in the particular case of phytophagous insects).
>
> **Host-plant adaptation:** heritable trait that confers a selective advantage on a particular host-plant.
>
> **Phylogenetic conservatism**: tendency for closely related species to be more similar than expected under Brownian motion evolution.
>
> **Phylogenetic tracking:** it occurs when a host-dependent species (here a phytophagous insect) diversifies and utilizes niches created by the speciation its hosts (here host-plants), this leads to parallel phylogenetic trees but no temporal congruence of speciation events.
>
> **Phytophagous insect:** an insect that feeds on any plant organ during whole or part of its life cycle, it excludes pollinators feeding on nectar and pollen but include pollinators that feed on developing seeds (*i.e.* seminiphagous insects).
>
> **Specialists:** species that use a narrow niche (a restricted set of resources).
>
> **Sister species/ sister clades:** the closest relatives of another given unit (species/ clade) in a phylogenetic tree.

**Conflict of interest disclosure**

The authors of this preprint declare that they have no financial conflict of interest with the content of this article. Emmanuelle Jousselin and Marianne Elias are recommenders for PCI Evol Biol. Marianne Elias is a member of the managing board of PCI Evol Biol.

## Author's contributions

EJ conceived the study, and extensively discussed the topic with ME. EJ wrote the first draft and both authors contributed to this and the previous versions of the manuscript.

## Acknowledgment

Version 4 of this preprint has been peer-reviewed and recommended by Peer Community In Evolutionary Biology (https://doi.org/10.24072/pci.evolbiol.100084). We thank Hervé Sauquet, Brian O'Meara and one anonymous reviewer for critically reading the manuscript and making very useful suggestions that helped improving its content.

## References

Agrawal, AA, M Fishbein, R Halitschke, AP Hastings, DL Rabosky, S Rasmann. 2009. Evidence for adaptive radiation from a phylogenetic study of plant defenses. Proceedings of the National Academy of Sciences of the United States of America 106:18067-18072.

Alfaro, ME, F Santini, C Brock, H Alamillo, A Dornburg, DL Rabosky, G Carnevale, LJ Harmon. 2009. Nine exceptional radiations plus high turnover explain species diversity in jawed vertebrates. Proceedings of the National Academy of Sciences of the United States of America 106:13410-13414.

Althoff, DM, KA Segraves, MTJ Johnson. 2014. Testing for coevolutionary diversification: linking pattern with process. Trends in Ecology & Evolution 29:82-89.

Althoff, DM, KA Segraves, CI Smith, J Leebens-Mack, O Pellmyr. 2012. Geographic isolation trumps coevolution as a driver of yucca and yucca moth diversification. Molecular Phylogenetics and Evolution 62:898-906.

Barraclough, TG, AP Vogler. 2000. Detecting the geographical pattern of speciation from species-level phylogenies. American Naturalist 155:419-434.

Beaulieu, JM, MJ Donoghue. 2013. Fruit evolution and diversification in campanulid angiosperms. Evolution 67:3132-3144.

Beaulieu, JM, BC O'Meara. 2016. Detecting Hidden Diversification Shifts in Models of Trait-Dependent Speciation and Extinction. Systematic Biology 65:583-601.

Becerra, JX. 1997. Insects on plants: macroevolutionary chemical trends in host use. Science 276:253-256.

Becerra, JX. 2003. Synchronous coadaptation in an ancient case of herbivory. Proceedings of the National Academy of Sciences of the United States of America 100:12804-12807.

Becerra, JX, DL Venable. 1999. Macroevolution of insect-plant associations: the relevance of host biogeography to host affiliation. Proceedings of the National Academy of Sciences of the United States of America 96:12626-12631.

Berlocher, SH. 2000. Radiation and divergence in the *Rhagoletis pomonella* species group: inferences from allozymes. Evolution 54:543-557.

Berry V, Chevenet F, Doyon JP, Jousselin E (2018) A geography-aware reconciliation method to investigate diversification patterns in host/parasite interactions. *Mol Ecol Resour* **18**, 1173-1184.

Blackman, RL, VF Eastop. 2006. Aphids on the World's Herbaceous Plants and Schrubs. London, UK: The Natural History Museum.

Bouchenak-Khelladi, Y, RE Onstein, YW Xing, O Schwery, HP Linder. 2015. On the complexity of triggering evolutionary radiations. New Phytologist 207:313-326.

Braga, MP, PR Guimarães, CW Wheat, S Nylin, N Janz. 2018. Unifying host-associated diversification processes using butterfly–plant networks. Nature Communications 9:5155.

Bramer, C, S Dobler, J Deckert, M Stemmer, G Petschenka. 2015. Na+/K+-ATPase resistance and cardenolide sequestration: basal adaptations to host plant toxins in the milkweed bugs (Hemiptera: Lygaeidae: Lygaeinae). Proceedings of the Royal Society B-Biological Sciences 282.

Brandle, M, S Knoll, S Eber, J Stadler, R Brandl. 2005. Flies on thistles: support for synchronous speciation? Biological Journal of the Linnean Society 84:775-783.

Brookes, DR, JP Hereward, LI Terry, GH Walter. 2015. Evolutionary dynamics of a cycad obligate pollination mutualism – Pattern and process in extant *Macrozamia* cycads and their specialist thrips pollinators. Molecular Phylogenetics and Evolution 93:83-93.

Brues, CT. 1924. The Specificity of Food-Plants in the Evolution of Phytophagous Insects. The American Naturalist 58:127-144.

Bush, GL. 1975. Sympatric Speciation in Phytophagous Parasitic Insects. In: PW Price, editor. Evolutionary Strategies of Parasitic Insects and Mites. Boston, MA: Springer US. p. 187-206.

Caillaud, MC, S Via. 2000. Specialized feeding behavior influences both ecological specialization and assortative mating in sympatric host races of pea aphids. American Naturalist 156:606-621.

Challis RJ, Kumar S, Dasmahapatra KKK, Jiggins CD, Blaxter M. 2016. Lepbase: the Lepidopteran genome database. *BioRxiv*, 056994.

Chan, KMA, BR Moore. 2002. Whole-tree methods for detecting differential diversification rates. Systematic Biology 51:855-865.

Conow, C, D Fielder, Y Ovadia, R Libeskind-Hadas. 2010. Jane: a new tool for the cophylogeny reconstruction problem. Algorithms for Molecular Biology 5.

Cruaud, A, N Ronsted, B Chantarasuwan, et al. 2012. An extreme case of plant-insect codiversification: figs and fig-pollinating wasps. Systematic Biology 61:1029-1047.

Davis, MP, PE Midford, W Maddison. 2013. Exploring power and parameter estimation of the BiSSE method for analyzing species diversification. BMC Evolutionary Biology 13:38.

de Castro, ÉCP, M Zagrobelny, MZ Cardoso, S Bak. 2018. The arms race between heliconiine butterflies and Passiflora plants – new insights on an ancient subject. Biological reviews 93:555-573.

de Vienne, DM, G Refregier, M Lopez-Villavicencio, A Tellier, ME Hood, T Giraud. 2013. Cospeciation vs host-shift speciation: methods for testing, evidence from natural associations and relation to coevolution. New Phytologist 198:347-385.

Doorenweerd, C, EJ van Nieukerken, SBJ Menken. 2015. A Global Phylogeny of Leafmining *Ectoedemia* Moths (Lepidoptera: Nepticulidae): Exploring Host Plant Family Shifts and Allopatry as Drivers of Speciation. Plos One 10.

Drès, M, J Mallet. 2002. Host races in plant-feeding insects and their importance in sympatric speciation. Philosophical Transactions of the Royal Society of London Series B-Biological Sciences 357:471-492.

Drury, JP, GF Grether, T Garland, Jr., H Morlon. 2017. An assessment of phylogenetic tools for analyzing the interplay between interspecific interactions and phenotypic evolution. Systematic Biology 67:413-427.

Ebel, ER, JM DaCosta, MD Sorenson, RI Hill, AD Briscoe, KR Willmott, SP Mullen. 2015. Rapid diversification associated with ecological specialization in Neotropical *Adelpha* butterflies. Molecular Ecology 24:2392-2405.

Edger, PP, HM Heidel-Fischer, M Bekaert, et al. 2015. The butterfly plant arms-race escalated by gene and genome duplications. Proceedings of the National Academy of Sciences of the United States of America 112:8362-8366.

Egan, SP, P Nosil, DJ Funk. 2008. Selection and genomic differentiation during ecological speciation: Isolating the contributions of host association via a comparative genome scan of *Neochlamisus bebbianae* leaf beetles. Evolution 62:1162-1181.

Ehrlich, PR, PH Raven. 1964. Butterflies and plants: a study in coevolution. Evolution 18:586-608.

Elias, M, R Faria, Z Gompert, A Hendry. 2012. Factors Influencing Progress toward Ecological Speciation. International Journal of Ecology 2012:7.

Elias, M, C Fontaine, FJF van Veen. 2013. Evolutionary history and ecological processes shape a local multilevel antagonistic network. Current Biology 23:1355-1359.

Endara, MJ, PD Coley, G Ghabash, JA Nicholls, KG Dexter, DA Donoso, GN Stone, RT Pennington, TA Kursar. 2017. Coevolutionary arms race versus host defense chase in a tropical herbivore-plant system. Proceedings of the National Academy of Sciences of the United States of America 114:E7499-E7505.

Etienne, RS, B Haegeman, T Stadler, T Aze, N Pearson Paul, A Purvis, B Phillimore Albert. 2012. Diversity-dependence brings molecular phylogenies closer to agreement with the fossil record. Proceedings of the Royal Society B: Biological Sciences 279:1300-1309.

Faith, DP, PS Cranston. 1991. Could a cladogram this short have arisen by chance alone? : on permutation for cladistic structure. Cladistics 7:1-28.

Farrell, BD, C Mitter. 1998. The timing of insect/plant diversification: might *Tetraopes* (Coleoptera: Cerambycidae) and *Asclepias* (Asclepiadaceae) have co- evolved? Biological Journal of the Linnean Society 63:553-577.

Fatouros, NE, C Broekgaarden, G Bukovinszkine'Kiss, JJA van Loon, R Mumm, ME Huigens, M Dicke, M Hilker. 2008. Male-derived butterfly anti-aphrodisiac mediates induced indirect plant defense. Proceedings of the National Academy of Sciences 105:10033-10038.

FitzJohn, RG. 2010. Quantitative Traits and Diversification. Systematic Biology 59:619-633.

Forbes, AA, SN Devine, AC Hippee, ES Tvedte, AKG Ward, HA Widmayer, CJ Wilson. 2017. Revisiting the particular role of host shifts in initiating insect speciation. Evolution 71:1126-1137.

Fordyce, JA. 2010. Host shifts and evolutionary radiations of butterflies. Proceedings of the Royal Society B: Biological Sciences 277:3735-3743.

Forister, ML, LA Dyer, MS Singer, JO Stireman, JT Lill. 2012. Revisiting the evolution of ecological specialization, with emphasis on insect-plant interactions. Ecology 93:981-991.

Funk, DJ, KE Filchak, JL Feder. 2002. Herbivorous insects: model systems for the comparative study of speciation ecology. Genetica 116:251-267.

Futuyma, DJ, AA Agrawal. 2009. Macroevolution and the biological diversity of plants and herbivores. Proceedings of the National Academy of Sciences 106:18054-18061.

García Morales M, BD Denno, GL Miller, *et al.* 2016. ScaleNet: a literature-based model of scale insect biology and systematics. Database 2016.

Godsoe, W, E Strand, CI Smith, JB Yoder, TC Esque, O Pellmyr. 2009. Divergence in an obligate mutualism is not explained by divergent climatic factors. New Phytologist 183:589-599.

Gómez-Zurita, J, T Hunt, F Kopliku, AP Vogler. 2007. Recalibrated tree of leaf beetles (Chrysomelidae) indicates independent diversification of angiosperms and their insect herbivores. Plos One 2:e360.

Goswami, A, PD Mannion, MJ Benton. 2016. Radiation and extinction: investigating clade dynamics in deep time. Biological Journal of the Linnean Society 118:6-12.

Gow, JL, CL Peichel, EB Taylor. 2007. Ecological selection against hybrids in natural populations of sympatric threespine sticklebacks. Journal of Evolutionary Biology 20:2173-2180.

Hamm, CA, JA Fordyce. 2015. Patterns of host plant utilization and diversification in the brush-footed butterflies. Evolution 69:589-601.

Hamm, CA, JA Fordyce. 2016. Greater host breadth still not associated with increased diversification rate in the Nymphalidae-A response to Janz et al. Evolution 70:1156-1160.

Hardy, NB. 2017. Do plant-eating insect lineages pass through phases of host-use generalism during speciation and host switching? Phylogenetic evidence. Evolution 71:2100-2109.

Hardy, NB, SP Otto. 2014. Specialization and generalization in the diversification of phytophagous insects: tests of the musical chairs and oscillation hypotheses. Proceedings of the Royal Society B-Biological Sciences 281.

Hardy, NB, DA Peterson, BB Normark. 2016. Nonadaptive radiation: Pervasive diet specialization by drift in scale insects? Evolution 70:2421-2428.

Hembry, DH, JB Yoder, KR Goodman. 2014. Coevolution and the Diversification of Life. American Naturalist 184:425-438.

Hernández-Vera G, Toševski I, Caldara R, Emerson BC. 2019. Evolution of host plant use and diversification in a species complex of parasitic weevils (Coleoptera: Curculionidae). *PeerJ* **7:** e6625.

Holman, J. 2009. Host Plant Catalog of Aphids: Palaearctic Region. Springer, Dordrecht.

Holt, RD. 1977. Predation, apparent competition, and the structure of prey communities. Theoretical Population Biology 12:197-229.

Hunt, T, J Bergsten, Z Levkanicova, et al. 2007. A comprehensive phylogeny of beetles reveals the evolutionary origins of a superradiation. Science 318:1913-1916.

Hsu YH, Cocroft RB, Snyder RL, Lin CP. 2018. You stay, but I Hop: Host shifting near and far co-dominated the evolution of Enchenopa treehoppers. Ecology and Evolution 8: 1954-1965.Peer Community In Evolutionary Biology    28

Ives, AR, HCJ Godfray. 2006. Phylogenetic analysis of trophic associations. American Naturalist 168:E1-E14.

Janz, N. 2011. Ehrlich and Raven revisited: mechanisms underlying codiversification of plants and enemies. In: DJ Futuyma, HB Shaffer, D Simberloff, editors. Annual Review of Ecology, Evolution, and Systematics. p. 71-89.

Janz, N, MP Braga, N Wahlberg, S Nylin. 2016. On oscillations and flutterings—A reply to Hamm and Fordyce. Evolution. 70:1150-1155.

Janz, N, K Nyblom, S Nylin. 2001. Evolutionary dynamics of host-plant specialization: A case study of the tribe Nymphalini. Evolution 55:783-796.

Janz, N, S Nylin. 2008. The oscillation hypothesis of host-plant range and speciation. In: D Tilman, editor. The evolutionary biology of herbivorous insects: specialization, speciation and radiation. University of california press p. 203-215.

Janz, N, S Nylin, N Wahlberg. 2006. Diversity begets diversity: host expansions and the diversification of plant-feeding insects. BMC Evolutionary Biology 6.

Jermy, T. 1976. Insect—Host-plant Relationship—Co-evolution or Sequential Evolution? The host-plant in relation to insect behaviour and reproduction: Springer. p. 109-113.

Jermy, T. 1984. Evolution of insect/ plant relationship. The American Naturalist 124:609-630.

Jermy, T. 1988. Can Predation Lead to Narrow Food Specialization in Phytophagous Insects? Ecology 69, 902-904.

Jordal, B, G M Hewitt. 2004. The Origin and Radiation of Macaronesian Beetles Breeding in *Euphorbia*: The Relative Importance of Multiple Data Partitions and Population Sampling. Systematic Biology 53:711-734.

Jousselin, E, A Cruaud, G Genson, F Chevenet, RG Foottit, AC d'acier. 2013. Is ecological speciation a major trend in aphids? Insights from a molecular phylogeny of the conifer-feeding genus *Cinara*. Frontiers in Zoology 10.

Jousselin, E, G Genson, A Coeur d'Acier. 2010. Evolutionary lability of a complex life cycle in the aphid genus *Brachycaudus*. BMC Evolutionary Biology 10.

Jousselin, E, JY Rasplus, F Kjellberg. 2003. Convergence and coevolution in a mutualism: Evidence from a molecular phylogeny of *Ficus*. Evolution 57:1255-1269.

Joy, JB, BJ Crespi. 2012. Island phytophagy: explaining the remarkable diversity of plant-feeding insects. Proceedings of the Royal Society B-Biological Sciences 279:3250-3255.

Kelley, ST, BD Farrell. 1998. Is specialization a dead end? The phylogeny of host use in *Dendroctonus* bark beetles (Scolytidae). Evolution 52:1731-1743.

Kergoat, GJ, FL Condamine, EFA Toussaint, C Capdevielle-Dulac, A-L Clamens, J Barbut, PZ Goldstein, B Le Ru. 2018. Opposite macroevolutionary responses to environmental changes in grasses and insects during the Neogene grassland expansion. Nature Communications 9:5089.

Kergoat, GJ, BP Le Ru, G Genson, C Cruaud, A Couloux, A Delobel. 2011. Phylogenetics, species boundaries and timing of resource tracking in a highly specialized group of seed beetles (Coleoptera: Chrysomelidae: Bruchinae). Molecular Phylogenetics and Evolution 59:746-760.

Kergoat, GJ, BP Le Ru, SE Sadeghi, M Tuda, CAM Reid, Z Gyorgy, G Genson, CS Ribeiro-Costa, A Delobel. 2015. Evolution of *Spermophagus* seed beetles (Coleoptera, Bruchinae, Amblycerini) indicates both synchronous and delayed colonizations of host plants. Molecular Phylogenetics and Evolution 89:91-103.

Labandeira, CC, ED Currano. 2013. The Fossil Record of Plant-Insect Dynamics. Annual Review of Earth and Planetary Sciences 41:287-311.

Labandeira, CC, DL Dilchet, DR Davis, DL Wagner. 1994. Ninety-seven million years of angiosperm-insect association: Paleobiological insights into the meaning of coevolution. Proceedings of the National Academy of Sciences of the United States of America 91:12278-12282.

Legendre, P, Y Desdevises, E Bazin. 2002. A statistical test for host parasite coevolution. Systematic Biology 51:217-234.

Leppanen, SA, E Altenhofer, AD Liston, T Nyman. 2012. Phylogenetics and evolution of host-plant use in leaf-mining sawflies (Hymenoptera: Tenthredinidae: Heterarthrinae). Molecular Phylogenetics and Evolution 64:331-341.

Leschen, RAB, TR Buckley. 2007. Multistate characters and diet shifts: Evolution of Erotylidae (Coleoptera). Systematic Biology 56:97-112.

Letsch H, Gottsberger B, Metzl C, Astrin J, Friedman ALL, McKenna DD, Fiedler K. 2018. Climate and host-plant associations shaped the evolution of ceutorhynch weevils throughout the Cenozoic. *Evolution* 72: 1815-1828.

Lin, Y-P, Dh Cook, Pj Gullan, Lg Cook. 2015. Does host-plant diversity explain species richness in insects? A test using Coccidae (Hemiptera). Ecological Entomology 40:299-306.

Linnen, CR, BD Farrell. 2010. A test of the sympatric host race formation hypothesis in *Neodiprion* (Hymenoptera: Diprionidae). Proceedings of the Royal Society B-Biological Sciences 277:3131-3138.

Lisa De-Silva, D, LL Mota, N Chazot, et al. 2017. North Andean origin and diversification of the largest ithomiine butterfly genus. Scientific Reports 7:45966.

Livshultz, T, E Kaltenegger, SCK Straub, K Weitemier, E Hirsch, K Koval, L Mema, A Liston. 2018. Evolution of pyrrolizidine alkaloid biosynthesis in Apocynaceae: revisiting the defence de-escalation hypothesis. The New phytologist 218:762-773.

Lopez-Vaamonde, C, N Wikstrom, C Labandeira, HCJ Godfray, SJ Goodman, JM Cook. 2006. Fossil-calibrated molecular phylogenies reveal that leaf-mining moths radiated millions of years after their host plants. Journal of Evolutionary Biology 19:1314-1326.

Loss-Oliveira, L, BO Aguiar, CG Schrago. 2012. Testing Synchrony in Historical Biogeography: The Case of New World Primates and Hystricognathi Rodents. Evolutionary Bioinformatics 8:127-137.

Maddison, WP, RG FitzJohn. 2015. The unsolved challenge to phylogenetic correlation tests for categorical characters. Systematic Biology 64:127-136.

Maddison, WP, PE Midford, SP Otto. 2007. Estimating a binary character's effect on speciation and extinction. Systematic Biology 56:701-710.

Manceau, M, A Lambert, H Morlon. 2016. A unifying comparative phylogenetic framework including traits coevolving across interacting lineages. Systematic Biology 66:551-568.

Marquis, RJ, D Salazar, C Baer, J Reinhardt, G Priest, K Barnett. 2016. Ode to Ehrlich and Raven or how herbivorous insects might drive plant speciation. Ecology 97:2939-2951.

Martínez-Aquino, A. 2016. Phylogenetic framework for coevolutionary studies: a compass for exploring jungles of tangled trees. Current Zoology 62:393-403

Matsubayashi, KW, I Ohshima, P Nosil. 2010. Ecological speciation in phytophagous insects. Entomologia Experimentalis et Applicata 134:1-27.

Matsuo, T. 2008. Rapid Evolution of Two Odorant-Binding Protein Genes, *Obp57* and *Obp57*, in the *Drosophila melanogaster* Species Group. Genetics 178:1061-1072.

McBride, CS, MC Singer. 2010. Field Studies Reveal Strong Postmating Isolation between Ecologically Divergent Butterfly Populations. Plos Biology 8:e1000529.

McKenna, DD, AS Sequeira, AE Marvaldi, BD Farrell. 2009. Temporal lags and overlap in the diversification of weevils and flowering plants. Proceedings of the National Academy of Sciences of the United States of America 106:7083-7088.

McKenna, DD, AL Wild, K Kanda, et al. 2015. The beetle tree of life reveals that Coleoptera survived end-Permian mass extinction to diversify during the Cretaceous terrestrial revolution. Systematic Entomology 40:835-880.

McLeish, MJ, JT Miller, LA Mound. 2013. Delayed colonisation of *Acacia* by thrips and the timing of host-conservatism and behavioural specialisation. BMC Evolutionary Biology 13.

Meseguer, AS, A Coeur d'acier, G Genson, E Jousselin. 2015. Unravelling the historical biogeography and diversification dynamics of a highly diverse conifer-feeding aphid genus. Journal of Biogeography 42:1482-1492.

Mitter, C, BD Farrell, BM Wiegman. 1988. The phylogenetic study of adaptive zones: has phytophagy promoted insect diversification? American Naturalist 132:107-128.

Moran, NA. 1992. The evolution of aphid life cycles. Annual Review of Entomology 37:321-348.

Mullen, SP, WK Savage, N Wahlberg, KR Willmott. 2011. Rapid diversification and not clade age explains high diversity in neotropical *Adelpha* butterflies. Proceedings of the Royal Society B-Biological Sciences 278:1777-1785.

Nakadai, R. 2017. Species diversity of herbivorous insects: a brief review to bridge the gap between theories focusing on the generation and maintenance of diversity. Ecological Research.

Nakadai, R, A Kawakita. 2016. Phylogenetic test of speciation by host shift in leaf cone moths (*Caloptilia*) feeding on maples (*Acer*). Ecology and Evolution 6: 4958–4970.

Nevado, B, GW Atchison, CE Hughes, DA Filatov. 2016. Widespread adaptive evolution during repeated evolutionary radiations in New World lupins. Nature Communications 7:12384.

Nosil, P. 2012. Ecological speciation. Oxford: Oxford University Press.

Nosil, P, BJ Crespi, CP Sandoval. 2002. Host-plant adaptation drives the parallel evolution of reproductive isolation. Nature 417:440-443.

Nosil, P, Z Gompert, TE Farkas, AA Comeault, JL Feder, CA Buerkle, TL Parchman. 2012. Genomic consequences of multiple speciation processes in a stick insect. Proceedings of the Royal Society B-Biological Sciences 279:5058-5065.

Nosil, P, AO Mooers. 2005. Testing hypotheses about ecological specialization using phylogenetic trees. Evolution 59:2256-2263.

Nylin, S, N Janz. 2009. Butterfly host plant range: an example of plasticity as a promoter of speciation? Evolutionary Ecology 23:137-146.

Nylin, S, J Slove, N Janz. 2014. Host plant utilization, host range oscillations and diversification in nymphalid butterflies: a phylogenetic investigation. Evolution 68:105-124.

Nylin, S, N Wahlberg. 2008. Does plasticity drive speciation? Host-plant shifts and diversification in nymphaline butterflies (Lepidoptera : Nymphalidae) during the tertiary. Biological Journal of the Linnean Society 94:115-130.

Nyman T, RE Onstein, D Silvestro, S Wutke, A Taeger, N Wahlberg, SM Blank, T Malm. 2019. The early wasp plucks the flower: disparate extant diversity of sawfly superfamilies (Hymenoptera: 'Symphyta') may reflect asynchronous switching to angiosperm hosts. Biological Journal of the Linnean Society 128: 1-19.

Nyman, T, HP Linder, C Pena, T Malm, N Wahlberg. 2012. Climate-driven diversity dynamics in plants and plant-feeding insects. Ecology Letters 15:889-898.

Nyman, T, V Vikberg, DR Smith, JL Boeve. 2010. How common is ecological speciation in plant-feeding insects? A 'Higher' Nematinae perspective. BMC Evolutionary Biology 10.

O'Meara, BC, JM Beaulieu. 2016. Past, future, and present of state-dependent models of diversification. American Journal of Botany 103:792-795.

Onstein, RE, WJ Baker, TLP Couvreur, S Faurby, J-C Svenning, WD Kissling. 2017. Frugivory-related traits promote speciation of tropical palms. Nature Ecology & Evolution 1:1903-1911.

Ouvrard D, Chalise P, Percy DM. 2015. Host-plant leaps versus host-plant shuffle: a global survey reveals contrasting patterns in an oligophagous insect group (Hemiptera, Psylloidea). Systematics and Biodiversity 13, 434-454.

Owen, CL, DC Marshall, KBR Hill, C Simon. 2017. How the aridification of australia structured the biogeography and influenced the diversification of a large lineage of australian cicadas. Systematic Biology 66:569-589.

Page, RDM. 1994. Parallel phylogenies: reconstructing the history of host-parasite assemblages. Cladistics 10:155-173.

Pagel, M. 1999. The maximum likelihood approach to reconstructing ancestral character states of discrete characters on phylogenies. Systematic Biology 48:612-622.

Peccoud, J, A Ollivier, M Plantegenest, JC Simon. 2009. A continuum of genetic divergence from sympatric host races to species in the pea aphid complex. Proceedings of the National Academy of Sciences of the United States of America 106:7495-7500.

Pellmyr, O. 2003. Yuccas, yucca moths, and coevolution: a review. Annals of the Missouri Botanical Garden 90:35-55.

Pena, C, M Espeland. 2015. Diversity dynamics in nymphalidae butterflies: effect of phylogenetic uncertainty on diversification rate shift estimates. Plos One 10.

Pena, C, N Wahlberg. 2008. Prehistorical climate change increased diversification of a group of butterflies. Biology letters 4:274-278.

Pena, C, H Witthauer, I Kleckova, Z Fric, N Wahlberg. 2015. Adaptive radiations in butterflies: evolutionary history of the genus Erebia (Nymphalidae: Satyrinae). Biological Journal of the Linnean Society 116:449-467.

Pennell, MW, LJ Harmon. 2013. An integrative view of phylogenetic comparative methods: connections to population genetics, community ecology, and paleobiology. In: TA Mousseau, CW Fox, editors. Year in Evolutionary Biology. p. 90-105.

Pitteloud, C, N Arrigo, T Suchan, et al. 2017. Climatic niche evolution is faster in sympatric than allopatric lineages of the butterfly genus *Pyrgus*. Proceedings of the Royal Society B-Biological Sciences 284.

Powell, THQ, AA Forbes, GR Hood, JL Feder. 2014. Ecological adaptation and reproductive isolation in sympatry: genetic and phenotypic evidence for native host races of *Rhagoletis pomonella*. Molecular Ecology 23:688-704.

Rabosky, DL. 2006. LASER: A maximum likelihood toolkit for detecting temporal shifts in diversification rates from molecular phylogenies. Evolutionary Bioinformatics 2:247-250.

Rabosky, DL, SC Donnellan, M Grundler, IJ Lovette. 2014. Analysis and visualization of complex macroevolutionary dynamics: an example from australian scincid lizards. Systematic Biology 63:610-627.

Rabosky, DL, EE Goldberg. 2015. Model inadequacy and mistaken inferences of trait-dependent speciation. Systematic Biology 64:340-355.

Rabosky, DL, IJ Lovette. 2008. Explosive evolutionary radiations: decreasing speciation or increasing extinction through time? Evolution 62:1866-1875.

Ramirez, W. 1974. Coevolution of *Ficus* and Agaonidae. Annals of the Missouri Botanical Garden 61:770-780.

Ree, RH, SA Smith. 2008. Maximum likelihood inference of geographic range evolution by dispersal, local extinction, and cladogenesis. Systematic Biology 57:4-14.

Roskam, JC. 1985. Evolutionary patterns in gall midge - host plant associations (Diptera, Cecidomyiidae). Tijdschrift voor Entomologie 128:193-213.

Rundle, HD, L Nagek, J Wenrick Boughman, D Schluter. 2000. Natural selection and parallel speciation in sympatric Sticklebacks. Science 287:306-308.

Rundle, HD, P Nosil. 2005. Ecological speciation. Ecology Letters 8:336-352.

Sahoo, RK, AD Warren, SC Collins, U Kodandaramaiah. 2017. Host-plant change and paleoclimatic events explain diversification shifts in skipper butterflies (Family: Hesperiidae). BMC Evolutionary Biology 17:174.

Sánchez-Gracia, A, FG Vieira, J Rozas. 2009. Molecular evolution of the major chemosensory gene families in insects. Heredity 103:208.

Sanchez-Guillen, RA, A Cordoba-Aguilar, B Hansson, J Ott, M Wellenreuther. 2016. Evolutionary consequences of climate-induced range shifts in insects. Biological reviews 91:1050-1064.

Schluter, D. 2000. The Ecology of Adaptive Radiation. Oxford University Press, New York.

Schluter, D. 2009. Evidence for Ecological Speciation and Its Alternative. Science 323:737-741.

Schwander, T, D Arbuthnott, R Gries, G Gries, P Nosil, BJ Crespi. 2013. Hydrocarbon divergence and reproductive isolation in *Timema* stick insects. BMC Evolutionary Biology 13.

Segar, ST, C Lopez-Vaamonde, JY Rasplus, JM Cook. 2012. The global phylogeny of the subfamily Sycoryctinae (Pteromalidae): Parasites of an obligate mutualism. Molecular Phylogenetics and Evolution 65:116-125.

Simonsen, TJ, N Wahlberg, AD Warren, FAH Sperling. 2010. The evolutionary history of *Boloria* (Lepidoptera: Nymphalidae): phylogeny, zoogeography and larval-foodplant relationships. Systematics and Biodiversity 8:513-529.

Singer, MS, JO Stireman. 2005. The tri-trophic niche concept and adaptive radiation of phytophagous insects. Ecology Letters 8:1247-1255.

Slove, J, N Janz. 2011. The relationship between diet breadth and geographic range size in the butterfly subfamily Nymphalinae - A Study of Global Scale. Plos One 6.

Smadja, C, RK Butlin. 2009. On the scent of speciation: the chemosensory system and its role in premating isolation. Heredity 102:77-97.

Smadja, CM, B Canback, R Vitalis, M Gautier, J Ferrari, JJ Zhou, RK Butlin. 2012. Large-scale candidate gene scan reveals the role of chemoreceptor genes in host plant specialization and speciation in the pea aphid. Evolution 66:2723-2738.

Stadler, T. 2013. Recovering speciation and extinction dynamics based on phylogenies. Journal of Evolutionary Biology 26:1203-1219.

Stadler, T, F Bokma. 2013. Estimating Speciation and Extinction Rates for Phylogenies of Higher Taxa. Systematic Biology 62:220-230.

Stone, GN, A Hernández-Vera -Lopez, JA Nicholls, E di Pierro, J Pujade-Villar, G Melika, JM Cook. 2009. Extreme Host Plant Conservatism During at Least 20 Million Years of Host Plant Pursuit by Oak Gallwasps. Evolution 63:854-869.

Suchan, T, N Alvarez. 2015. Fifty years after Ehrlich and Raven, is there support for plant-insect coevolution as a major driver of species diversification? Entomologia Experimentalis et Applicata 157:98-112.

Thébault, E, C Fontaine. 2010. Stability of ecological communities and the architecture of mutualistic and trophic networks. Science 329:853-856.

Thompson, JN. 1989. Concepts of coevolution. Trends in Ecology and Evolution 4:179-183.

Vea, IM, DA Grimaldi. 2016. Putting scales into evolutionary time: the divergence of major scale insect lineages (Hemiptera) predates the radiation of modern angiosperm hosts. Scientific Reports 6:23487.

Via, S, AC Bouck, S Skillman. 2000. Reproductive isolation between divergent races of pea aphids on two hosts. II. Selection against migrants and hybrids in the parental environments. Evolution 54:1626-1637.

Vieira, FG, J Rozas. 2011. Comparative genomics of the odorant-binding and chemosensory protein gene families across the arthropoda: origin and evolutionary history of the chemosensory system. Genome Biology and Evolution 3:476-490.

Wahlberg, N. 2001. The phylogenetics and biochemistry of host-plant-specialization in Melitaeninae butterflies (Lepidoptera: Nymphalidae). Evolution 55:522-537.

Wahlberg, N, CW Wheat, C Pena. 2013. Timing and Patterns in the Taxonomic Diversification of Lepidoptera (Butterflies and Moths). Plos One 8.

Walsh, B. 1864. On phytophagic varieties and phytophagis species. Proceedings of the Entomological Society, Philadelphia:403-430.

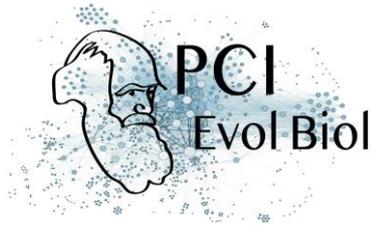

Wang, HS, JD Holloway, N Janz, MP Braga, N Wahlberg, M Wang, S Nylin. 2017. Polyphagy and diversification in tussock moths: support for the oscillation hypothesis from extreme generalists. Ecology and Evolution 7:7975-7986.

Weiblen, GD. 2004. Correlated evolution in fig pollination. Systematic Biology 53:128-139.

Weingartner, E, N Wahlberg, S Nylin. 2006. Dynamics of host plant use and species diversity in Polygonia butterflies (Nymphalidae). Journal of Evolutionary Biology 19:483-491.

Wheat, CW, H Vogel, U Wittstock, MF Braby, D Underwood, T Mitchell-Olds. 2007. The genetic basis of a plant-insect coevolutionary key innovation. Proceedings of the National Academy of Sciences of the United States of America 104:20427-20431.

Wiebes, JT. 1979. Co-evolution of figs and their insect pollinators. Annual Review of Ecology and Systematics 10:1-12.

Wiens, JJ, RT Lapoint, NK Whiteman. 2015. Herbivory increases diversification across insect clades. Nature Communications 6.

Wilson, JS, ML Forister, LA Dyer, et al. 2012. Host conservatism, host shifts and diversification across three trophic levels in two Neotropical forests. Journal of Evolutionary Biology 25:532-546.

Winkler, I, SJ Scheffer, ML Lewis, KJ Ottens, AP Rasmussen, GA Gomes-Costa, LM Huerto Santillan, MA Condon, AA Forbes. 2018. Anatomy of a Neotropical insect radiation. BMC Evolutionary Biology 18:30.

Winkler, IS, C Mitter. 2008. The phylogenetic dimension of insect-plant interactions: a review of recent evidence. In: K Tilmon, editor. Specialization, Speciation and Radiation: the Evolutionary Biology of Herbivorous insects: University of California Press p. 240-263.

Winkler, IS, C Mitter, SJ Scheffer. 2009. Repeated climate-linked host shifts have promoted diversification in a temperate clade of leaf-mining flies. Proceedings of the National Academy of Sciences of the United States of America 106:18103-18108.

Winter S, Friedman ALL, Astrin JJ, Gottsberger B, Letsch H. 2017. Timing and host plant associations in the evolution of the weevil tribe Apionini (Apioninae, Brentidae, Curculionoidea, Coleoptera) indicate an ancient co-diversification pattern of beetles and flowering plants. Molecular Phylogenetics and Evolution 107: 179-190.

Yokoyama, J. 1995. Insect-plant coevolution and cospeciation. In: R Arai, M Kato, Y Doi, editors. Biodiversity and Evolution. Tokyo. p. 115-130.